\documentclass[preprint]{aastex}

\title{THE RISE TIME OF TYPE~Ia SUPERNOVAE FROM THE SUPERNOVA LEGACY 
 SURVEY$^1$}
\shorttitle{RISE TIME OF SNe~Ia FROM SNLS}
\shortauthors{Conley \etal}
\author{
 A.~Conley\altaffilmark{2},
 D.~A.~Howell\altaffilmark{2},
 A.~Howes\altaffilmark{2},
 M.~Sullivan\altaffilmark{2}, 
 P.~Astier\altaffilmark{3},
 D.~Balam\altaffilmark{4}, 
 S.~Basa\altaffilmark{5}, 
 R.~G.~Carlberg\altaffilmark{2},
 D.~Fouchez\altaffilmark{6}, 
 J.~Guy\altaffilmark{3}, 
 I.~Hook\altaffilmark{7}, 
 J.~D.~Neill\altaffilmark{4}, 
 R.~Pain\altaffilmark{3}, 
 K.~Perrett\altaffilmark{2},
 C.~J.~Pritchet\altaffilmark{4}, 
 N.~Regnault\altaffilmark{3}, 
 J.~Rich\altaffilmark{8}, 
 R.~Taillet\altaffilmark{9,3}, 
 E.~Aubourg\altaffilmark{10,8}, 
 J.~Bronder\altaffilmark{7}, 
 R.~S.~Ellis\altaffilmark{11}, 
 S.~Fabbro\altaffilmark{12},
 M.~Filiol\altaffilmark{5}, 
 D.~Le~Borgne\altaffilmark{8}
 N.~Palanque-Delabrouille\altaffilmark{8}, 
 S.~Perlmutter\altaffilmark{13},
 P.~Ripoche\altaffilmark{6}
}

\email{ conley@astro.utoronto.ca }
\altaffiltext{1}{
 Based on observations obtained with MegaPrime/MegaCam, a joint project
 of CFHT and CEA/DAPNIA, at the Canada-France-Hawaii Telescope (CFHT)
 which is operated by the National Research Council (NRC) of Canada,
 the Institut National des Sciences de l'Univers of the Centre National
 de la Recherche Scientifique (CNRS) of France, and the University of
 Hawaii. This work is based in part on data products produced at the
 Canadian Astronomy Data Centre as part of the Canada-France-Hawaii
 Telescope Legacy Survey, a collaborative project of NRC and CNRS.
}
\altaffiltext{2}{Department of Astronomy and Astrophysics, University
  of Toronto, 50 St. George Street, Toronto, ON M5S 3H4, Canada}
\altaffiltext{3}{LPNHE, CNRS-IN2P3 and University of Paris
  VI \& VII, 75005 Paris, France}
\altaffiltext{4}{Department of Physics and Astronomy, University of
  Victoria, PO Box 3055, Victoria, BC V8W 3P6, Canada}
\altaffiltext{5}{LAM CNRS,
  BP8, Traverse du Siphon, 13376 Marseille Cedex 12, France}
\altaffiltext{6}{CPPM, CNRS-IN2P3 and University Aix Marseille II,
  Case 907, 13288 Marseille Cedex 9, France}
\altaffiltext{7}{University of Oxford Astrophysics, Denys Wilkinson
  Building, Keble Road, Oxford OX1 3RH, UK}
\altaffiltext{8}{DAPNIA/Service d'Astrophysique, CEA-Saclay, 91191 
  Gif-sur-Yvette Cedex, France}
\altaffiltext{9}{Universit\'e de Savoie, 73000 Chambery, France}
\altaffiltext{10}{APC, 11 Pl. M.
  Berthelot, 75231 Paris Cedex 5, France}
\altaffiltext{11}{California Institute of Technology, E. California
  Blvd., Pasadena, CA 91125, USA}
\altaffiltext{12}{CENTRA - Centro Multidisciplinar de Astrof\'{\i}sica,
  IST, Avenida Rovisco Pais, 1049 Lisbon, Portugal}
\altaffiltext{13}{Lawrence Berkeley National Laboratory, 1 Cyclotron
  Rd., Berkeley, CA 94720, USA}

\newcommand{\chisq}{\ensuremath{\chi^2}}
\newcommand{\etal}{et~al.}
\newcommand{\kcorr}{\ensuremath{K}-correction}
\newcommand{\kcorrs}{\ensuremath{K}-corrections}
\newcommand{\sg}{\ensuremath{\sigma}}
\newcommand{\tr}{\ensuremath{\tau_r}}
\newcommand{\gp}{\ensuremath{g^{\prime}}}
\newcommand{\rp}{\ensuremath{r^{\prime}}}
\newcommand{\ip}{\ensuremath{i^{\prime}}}
\newcommand{\zp}{\ensuremath{z^{\prime}}}

\begin{document}

\begin{abstract}
We compare the rise times of nearby and distant Type~Ia supernovae (SNe~Ia)
as a test for evolution using 73 high-redshift spectroscopically-confirmed 
SNe~Ia from the first 2 years of the 5 year Supernova Legacy Survey 
(SNLS) and published observations of nearby SNe.  Because of the ``rolling'' 
search nature of the SNLS, our measurement is approximately 6 times more 
precise than previous studies, allowing for a more sensitive test of 
evolution between nearby and distant SNe.  Adopting a simple
$t^2$ early-time model (as in previous studies), we find that the 
rest-frame $B$ rise times for a fiducial SN~Ia at high and low redshift 
are consistent, with values $19.10^{+0.18}_{-0.17}\left(\mbox{stat}\right) 
\pm 0.2 \left(\mbox{syst}\right)$ and $19.58^{+0.22}_{-0.19}$ days, 
respectively;  the statistical significance of this difference is only 
1.4 \sg . The errors represent the uncertainty in the mean rather than any
variation between individual SNe.  We also compare subsets of our 
high-redshift data set based on decline rate, host galaxy star formation 
rate, and redshift, finding no substantive evidence for any subsample 
dependence.
\end{abstract}


\section{INTRODUCTION}
\nobreak
Type~Ia supernovae (SNe~Ia) have come to play a critical role in attempts 
to pin down the cosmological parameters.  Their utility arises because they 
appear to constitute a class of high-quality standardizable candles.
Considerable effort has been devoted by many groups to testing
this assumption.  Perhaps the most pernicious concern is that
the properties of SNe~Ia have evolved between the current epoch and
$z \sim 0.3-1$, which characterizes most current ``distant'' supernova samples.
One test for evolutionary effects is to compare the
rise time, the time from explosion until maximum luminosity, of the nearby
and distant samples.  The early light-curve behavior should be
influenced by the amount of $^{56}\mbox{Ni}$ synthesized in the explosion,
as well as the opacity of the ejecta \citep{Shigeyama:92,Branch:92,
Khokhlov:93,Vacca:96,Hoflich:93,Hoflich:98,Dominguez:01}.  
Changes in either of these, for example, due to changing progenitor 
metallicity with redshift, could affect the use of SNe~Ia as standard 
candles.  An evolutionary effect of 0.2 mag to $z=0.5$ would 
nullify the SN~Ia evidence that the Universe is accelerating,
and measuring $w$ to 10\% requires that any effect be smaller than 0.04 mag.
There are other routes to studying evolution with redshift: comparing
SNe~Ia in different host galaxy environments \citep{Hamuy:00,Sullivan:03,
Gallagher:05}, and via detailed spectroscopic studies (Hook \etal\ 2005, 
Blondin \etal\ 2006, Bronder \etal\ 2006, in preparation).  Neither 
approach has turned up any evidence of evolution.

The rise-time has implications for SNe~Ia explosion models.  Varying the rise 
time from 20 to 16 days at a fixed peak luminosity changes the implied 
amount of $^{56}$Ni synthesized in the explosion by $-10\%$ 
\citep{Contardo:00}.  Models of single white dwarf progenitor systems 
generally predict rise times in the range 13--19 days, while systems 
involving two white dwarfs allow for longer rise times because interaction 
with the disk of unaccreted material from the disrupted companion slows the 
diffusion of photons \citep{Hoflich:96, Hoflich:02}.

The determination of the rise times has been the subject of some
dispute. Historically, early-time, well-calibrated photometry of SNe~Ia has
been quite difficult to obtain.  Some of the earliest studies of SNe~Ia 
light-curve shapes examined the rise time \citep{Pskovskii:84}.  
\citet[hereafter R99]{Riess:99a} presented 
early observations of a set of nearby SNe~Ia detected 10--18 days before
maximum $B$ luminosity.  This data set was constructed primarily from 
unfiltered early detections, some from amateur observers.  They 
transformed these observations
to standard pass bands (in particular, $B$) using models of early-time 
SN~Ia spectra and colors.  They then considered various late-time light-curve 
parameterizations, concluding that the rise time for a fiducial SN~Ia
was $19.5 \pm 0.2$ days.  \citet{Riess:99b} compared this number
with the preliminary analysis of \citet[hereafter G01]{Goldhaber:01} for 
distant SNe~Ia and concluded that the 
rise times differed by $2.5 \pm 0.4$ days, a significance of 
$\sim 6\ \sg$, and a clear signature of evolution.  

\citet[hereafter AKN00]{AKN:00} argued that this comparison was based on 
analyses that had ignored the significant correlations between the 
light-curve parameters, and that taking these properly into account
increased the errors on the rise time to $\pm 1.2$ days for the distant 
sample. They concluded that the significance of the difference was 
closer to $1.5\ \sg$.

Our purpose here is not to revisit this controversy; rather, we
present a new, significantly more precise ($\sim 6$ times) measurement 
of the high-redshift rise time and compare it with the value for
nearby SNe.   This has particular relevance because current SN
projects place more stringent requirements on the standard-candle assumption.  
This paper presents measurements of the rise time from one such survey, 
the Supernova Legacy Survey (SNLS,\citet{Astier:06}).  The design of this 
(and some other modern surveys) results in a dramatic increase in the amount 
of early-time photometry available when compared with previous generations of 
surveys, as detailed in \S \ref{sec:data}.

We first describe the basic problem (\S \ref{sec:parameterization}), 
then present the data on which our measurement is based (\S \ref{sec:data}), 
followed by a description of our analysis procedures (\S \ref{sec:analysis}).
Finally, we compare our measurements against nearby SNe~Ia, as well as
between different subsets of our data (\S \ref{sec:results}).

\section{RISE-TIME PARAMETERIZATION}
\label{sec:parameterization}
\nobreak
In order to measure the rise time of our sample we require a mechanism for 
combining data from different SNe~Ia, correcting for the differences in 
light-curve shape and peak flux, and a model for the early-time flux as a 
function of time. Considerable effort and ingenuity have been devoted to 
developing techniques for parameterizing SNe~Ia light curves near and after
maximum light (Riess \etal\ 1996, G01, Guy \etal\ 2005).  Here we follow the 
stretch method as described in, for example, G01.  The flux as a function of 
time after the rise-time region is represented as
\begin{equation}
 f\left(t\right) = f_0\ \psi\left( \tau \right),
\end{equation}
where $f_0$ is the flux at maximum, and $\psi$ is some normalized flux template
appropriate to the passband under consideration. $\tau$ is
the effective date defined by 
$\tau = \left(t - t_{\mbox{max}}\right)/ s \left(1+z\right)$,
where $t_{\mbox{max}}$ is the date of maximum flux in some arbitrary filter
(usually $B$), $s$ is the stretch, and $z$ is the redshift.  
Conventionally, $s=1$ is defined to represent an average SN~Ia.
Once $s$, $f_0$, and $t_{\mbox{max}}$ are fited to the data, we 
can combine data from different SNe~Ia by
converting from observed epoch $t$ to $\tau$ and dividing by $f_0$ to 
normalize the flux values relative to each other.  We must also apply a \kcorr\
so that the resulting data are all expressed in the same rest-frame filter.
Note that different observations from the same SN are correlated by 
this procedure, and that in addition estimates of the light-curve fit 
parameters are generally quite strongly correlated.
We must take both into account in our analysis.
For our purposes we limit the fit to this model to the ``core'' light curve
between $-10 < \tau < 35$ days.  The lower limit arises because we fit the 
rise-time model in this range, and we want to prevent the fit from suppressing
unusual rise-time behavior.  The upper limit arises because after this 
effective epoch the SNe~Ia enter the so-called nebular phase, in which
the stretch prescription no longer works (G01).

In the rise-time region ($\tau < -10$), we follow earlier work 
(R99, AKN00, G01) in making use of a simple quadratic model 
\begin{equation}
 f \left( t \right) = \alpha \left( \tau + \tr \right)^2
 \label{eqn:risetime}
\end{equation}
for $\tau > -\tr$, and 0 at earlier times. \tr\ represents the
rise time for the ``fiducial'' $s=1$ template defined by $\psi$.
Because the template can vary from analysis to analysis, it is critical 
that any comparison between the nearby and distant
samples use the same one.  For our purposes, $\alpha$ is purely a 
nuisance parameter.  This approach implicitly assumes that stretch continues
to work well at early times, as shown by G01 and R99.  Our data set is
not particularly well suited to investigating more complicated relations 
between the rise time and stretch, although we have slightly more to 
say about this in \S \ref{sec:results}.  Note that we do not require continuity
between the fit to the rise-time region and the core light curve.

Following R99, the quadratic form can be
motivated by simple physical arguments:  at early times SNe~Ia should
be hot enough that the standard $BV$ pass-bands are in the Rayleigh-Jeans
tail of the spectral energy distribution (SED), so 
$f \propto r^2 T \propto v^2 \left(\tau + \tr \right)^2 T$
where $v$ is the velocity and $T$ the temperature.  This, coupled with 
the fact that the measured colors and velocities do not change rapidly 
compared to the time since explosion in this regime, 
suggests a $\left(\tau + \tr \right)^2$ behavior.  Ultimately, 
one of our goals is to test whether this model adequately describes 
the data.  However, it has some clear limitations.  In particular, one 
could always change the physical rise time in a fashion we could not detect by
adding an initial period of non-$t^2$ behavior at very low, and
hence undetectable, luminosities.

Because much of the R99 data comes from unfiltered, very broad band
observations, it can only be mapped to a single rest-frame
filter independently.  Hence, we are not yet in a position to measure the
$B$ and $V$ rise times separately for the nearby sample, and thus have
restricted this analysis to rest-frame $B$.  

\section{DATA}
\nobreak
\label{sec:data}
The SNLS\footnote{\url{http://www.cfht.hawaii.edu/SNLS/}} relies
on data from the deep component of the 5 year CFHT Legacy Survey, using
the square-degree MegaCam imager on the Canada-France-Hawaii Telescope 
\citep{Boulande:03}.  Repeat \gp \rp \ip \zp\ imaging is performed on 
four fields every three to four nights of dark time.  SNLS is a ``rolling'' 
search, in which the 
same fields are searched repeatedly for variable objects.  Each observation 
acts both as a potential discovery image for new SNe and as follow-up for 
candidates already discovered in the same field, 
allowing a considerable multiplex advantage.  The primary 
goal of this program is to measure the average equation-of-state parameter
$\left<w\right>$ of the dark energy to 5\% using $\sim 500-700$
high-redshift SNe~Ia (see \citep{Astier:06} for cosmological results from the 
first year of data).

The pioneering high-redshift SN programs \citep{Riess:98, Perlmutter:99} 
typically had a gap between the initial reference and the follow-up image 
used to discover new SNe of approximately 1 month.  This was designed to catch 
SNe around maximum luminosity and allow follow-up observations during
dark time, but it resulted in large gaps in the early-time coverage.  In fact, 
the majority of SNe~Ia discovered in this fashion 
have no early-time detections.  In a fraction of cases, however, SNe 
were discovered after maximum, or the redshift and light-curve shape were
such that one or two early data points were obtained.  This allowed G01 to 
measure the rise time at $z \approx 0.5$ to be $18.3 \pm 1.2$ days 
(statistical errors only).  In contrast, in a rolling search the gap 
between images is typically a few observer-frame days, and it is usually 
possible to go back to previous images
after the SN is discovered and measure the early-time flux.  Furthermore, 
the gaps in the light curve due to the lack of observations during
bright time are essentially uncorrelated with the light-curve phase.

In this paper we consider only SNe~Ia with spectroscopic type 
confirmation.  SN types were determined using observations with
the Gemini \citep{Howell:05}, ESO-Very Large Telescope (S.\ Basa \etal\ 2006, 
in preparation) and the W.M.\ Keck telescopes (R.\ S.\ Ellis \etal\ 2006, 
in preparation).  We make use of data from the first 2 years of SNLS 
and only consider SNe at redshifts below 0.88.  Above this redshift, 
our \ip\ filter maps most closely to rest-frame $U$ for a SN~Ia at maximum, and
so to measure the $B$ rise time there we depend heavily on our \zp\ 
observations.  These have a considerably lower signal to noise ratio because
the efficiency of our CCDs falls off at these wavelengths.
Furthermore, the \zp\ data
suffers from considerable fringing because the CCDs are thinned, which
results in additional scatter and calibration uncertainties.  Finally,
the \zp\ cadence is not as frequent as in the other filters.  The
resulting light-curve fits would be dominated by the rest-frame $U$-band
data, while we seek to measure the rise time directly in the $B$ band.
While it would certainly be possible to include these data, 
even with this restriction the systematic error is already dominant
in our rise-time determination.  In future SNLS data sets the \zp\
situation will improve, both because of changes to the fringe processing
and because more \zp\ observations are now obtained in each dark-time
cycle.

For our rise-time sample we require at least one rest-frame $B$ observation 
in the rise-time region ($-30 < \tau < -10$ days) and one near peak 
luminosity ($-6 < \tau < 6$ days)
to ensure that different light curves can be accurately normalized to each 
other using $f_0$.  We also require at least one rest-frame $U$, $B$, or $V$ 
observation between 10 and 35 effective days in order to ensure that the 
stretch is well determined.  These requirements eliminate roughly half of 
our SNe~Ia; the combination of our redshift range and the lunar cycle 
results in many SNe with data in the rise-time region not having observations 
near peak. However, the fraction of usable SNe for this purpose still far 
exceeds that possible with earlier data sets.  The requirements also mildly
select against the lower redshift portion of our SN sample, so the
median redshift of our rise-time sample is slightly higher than our
overall sample.  This results in a sample of 73 spectroscopically confirmed 
SNe~Ia in the desired redshift range.  We note that these requirements are 
far more stringent than those necessary for a cosmological analysis.

SNLS has two independent photometric pipelines, one based in Canada
and the other in France; this paper makes use of photometry from the
Canadian pipeline.  The details of this pipeline, as well as the resulting
photometry, will be presented elsewhere.  Briefly, photometry is performed
using a non-parametric PSF fit to subtracted images, for which the PSF is
derived from a set of `PSF stars' shared across all images of the same
field.  The SN images are never re-sampled; rather, the reference SN-free
images are transformed and PSF-matched to the data images.  As 
in \citet{Astier:06}, we rescale our photometric errors by $\sim +25\%$ to 
take into account the correlations in the reference image induced by 
these transformations.

It is obviously critical that the rise times of the nearby
and distant samples be computed using the same method.  R99
provides nearby photometry for epochs $\tau \lesssim -10$, so we supplement
this with data from a number of other sources in order to allow a consistent
determination of $f_0$ and $s$.  The sources are
provided in table~\ref{tbl:photsource}.  There are some potential complications
with combining the data sets.  In R99, the early photometry is given as
magnitude relative to peak as a function of epoch
relative to peak without specifying the peak magnitude or the date
of maximum precisely.  For most of the SNe this is not a serious 
problem,  as the light curves around peak are sufficiently well sampled
that both quantities can be determined relatively unambiguously.  For some
of the SNe, R99 used published early-time photometry, which allows the data 
to be tied together exactly.  However, we experienced problems with four
SNe, which we excluded from the sample.  First, we were
unable to locate any published photometry for SN1996by.  
For SN1996bv and SN1998ef the published photometry either has poor
coverage or is of sufficiently low quality that we cannot tightly
constrain the peak and date of maximum.
In addition to these three, SN1996bo has only $V$-band observations 
available in the rise-time region, so we have excluded it from our $B$ 
rise-time determination.  We note that the excluded SNe would add relatively 
little weight to the fit, as each SN only has 1 or 2 observations in the 
rise-time region.

To the R99 sample we add two more recent nearby SNe~Ia with good early 
coverage: SN 2001el and SN 2002bo.  The latter, in particular, adds 
considerably to the data sample.  We then have eight nearby SNe~Ia: 
SNe 1990N, 1994D, 1997bq, 1998aq, 1998bu, 1998dh, 2001el, and 2002bo.  

\begin{deluxetable}{llr}
\tablewidth{0pt}
\tablecaption{Nearby SNe~Ia used in the rise-time measurement
 \label{tbl:photsource} }
\tablehead{
 \colhead{SN} & \colhead{ $z$ } & \colhead{ Reference } 
}
\startdata
1990N  & 0.0034 & 1,2 \\
1994D  & 0.0015 & 1,3 \\
1997bq & 0.0094 & 1,4 \\
1998aq & 0.0037 & 1,5 \\
1998bu & 0.0030 & 1,4 \\
1998dh & 0.0089 & 1,4 \\
2001el & 0.0039 & 6   \\
2002bo & 0.0042 & 7,8 \\
\enddata
\tablerefs{ (1) R99, (2) \citet{Lira:98}, (3) \citet{Richmond:95},
 (4) \citet{Jha:06}, (5) \citet{Riess:05}, (6) \citet{Krisciunas:03},
 (7) \citet{Benetti:04}, (8) \citet{Krisciunas:04}}
\end{deluxetable}

\section{ANALYSIS}
\nobreak
\label{sec:analysis}
The primary complication we face in determining the rise time is in handling
correlations.  Particularly at high redshift, many SNe~Ia have multiple
observations in the rise-time region.  In order to combine observations
from different SNe~Ia the data must be flux normalized, and the
epochs must be shifted to reflect the date of maximum and the timescale
divided by the stretch and $1+z$.  The result is a set of data points with
significant correlations in both dimensions (time and flux).  Furthermore, 
the parameters determining the amount of correlation (the light-curve-fit 
parameters) are highly correlated themselves.  Directly addressing this 
situation through the covariance matrix of the early-time data is not
entirely trivial.  We have chosen a different approach, which is to use a 
Monte Carlo technique to handle the correlations.  We find that, for our
data sample, their effects are comparable in size to the contribution from 
the measurement uncertainties of the data points.  In other words, an 
analysis that ignored these correlations would underestimate the final 
error by approximately $\sqrt{2}$.  Worse, the resulting estimate for $\tr$ 
would be incorrect by several tenths of a day.

In order to fit the rise-time data, we need to know the light-curve
fit parameters (stretch, date of maximum, $f_0$).  We determine these
using the light-curve template of \citet{Knop:03} and a time series of 
spectral templates descended from those of \citet{Nugent:02}.  In our fits 
the flux scales are allowed to float independently in all filters, with 
only the stretch and date of maximum held fixed between different filters (the 
difference between the date of maximum in different rest-frame filters 
is set by our template).  Our fitting procedure predicts the SED of the 
SN on each observed epoch in physical units, which can then be converted 
to the $B$ flux.  We do not include data bluer than the rest-frame $U$ band 
in our fits, since there are very few observations of nearby SNe~Ia to 
constrain our model SEDs in this region.  This means that the \gp\ filter
is not used at redshifts above 0.4, which constitutes the majority of our
sample.  The outputs of this procedure are best-fit values for each of
the light-curve parameters as well as their correlations.

With the light-curve parameters in hand, we combine the data as described
in \S \ref{sec:parameterization}, applying the \kcorr\ based on the
model predicted SED.  We then fit the rise-time model defined in
equation~\ref{eqn:risetime} to the rest-frame $B$ fluxes in the range 
$-30 < \tau < -10$ days.  We minimize the \chisq\ of the model with respect to 
the data, fitting to the individual observations.  For the SNLS data there
are usually multiple observations of each SN in each filter on each epoch, 
so we can reject outliers due to, for example, unidentified cosmic rays by 
removing data points that disagree by more than 3.5 \sg\ with other 
observations on the same night.  This does not catch all outliers because 
for some nights there are only one or two calibratable observations in a 
given filter due to weather or other issues.  We therefore also apply a 
3.5 \sg\ outlier cut with respect to the model fit in an iterative fashion.  
This raises the possibility that interestingly discrepant SNe could be 
removed from the sample,
but we find that relaxing or removing this cut has no effect on the final 
answer except to increase the \chisq\ of the fit.  Furthermore,
a SN-by-SN investigation of data points that are removed by this cut
gives no convincing examples of unusual rise-time behavior.  As for
R99 and AKN00, varying the upper limit of the data included in our
parabolic fit ($\tau = -10$) has little effect on our results as long 
as it is earlier than $\tau \lesssim -8$ days.

This, however, does not address the question of how to handle the
induced correlations in the rise-time region.  One approach would be to 
use the error properties of the individual photometry points to generate a 
random realization of our data sample, fit the core light curve for each SN 
as described above to determine the light-curve parameters, and then
use these to perform the fit in the rise-time region to the combined
data.  After repeating this many times, the resulting 
distribution of \tr\ could be used to find the rise time and its associated 
errors. A useful simplification arises because, for the purposes of the
rise-time fit, the only part of the core light curve we are concerned with
are the light-curve fit parameters, and so we can work in this space instead 
of directly with the photometry to handle the correlations.  
The contribution to the final error on \tr\ 
is then split into two terms: that arising from the correlations between 
different points induced by the flux normalization and conversion to $\tau$,
and that from the random photometric noise of each measurement.

To handle the first term, we use the covariance matrix
between the light-curve fit parameters for each SN to randomly generate a 
large number of realizations (typically 2000) of those parameters for 
each SN in our sample using standard techniques \citep{James:75}.
Changing the light-curve fit parameters changes the model SED for each data 
point, and so the \kcorrs\ must be re-calculated and applied with each 
realization.  For each of these sets we fit the rise-time region to
determine specific values for \tr\ and $\alpha$, and then combine the 
results from all of the realizations to get the final values and their 
associated errors. The cost of this approach is that we cannot associate a 
simple \chisq\ statistic with our overall fit.  However, fits to individual
realizations generally result in acceptable \chisq\ values (for the main
result presented below the \chisq\ is 1860 for 1409 degrees of freedom),
especially considering that the correlations are not included in these
numbers.  This indicates that the quadratic rise-time model
is a good representation of our data.  The correlations affect both the 
fit value and errors for \tr\ significantly.  

We can then include the term from the variances of the individual points
by randomly re-sampling the rise time and $\alpha$ values using
the error reported for each individual rise-time fit. The resulting errors 
on \tr\ can be found by finding the ranges that contain the desired 
fraction of the total probability around the mean value.  We use limits 
that are symmetric in probability space around the mean.  The 
distributions have somewhat non-Gaussian tails, so the 2 \sg\ errors are 
generally not exactly twice the size of the the 1 \sg\ errors, etc.

\section{RESULTS}
\begin{figure}
\plotone{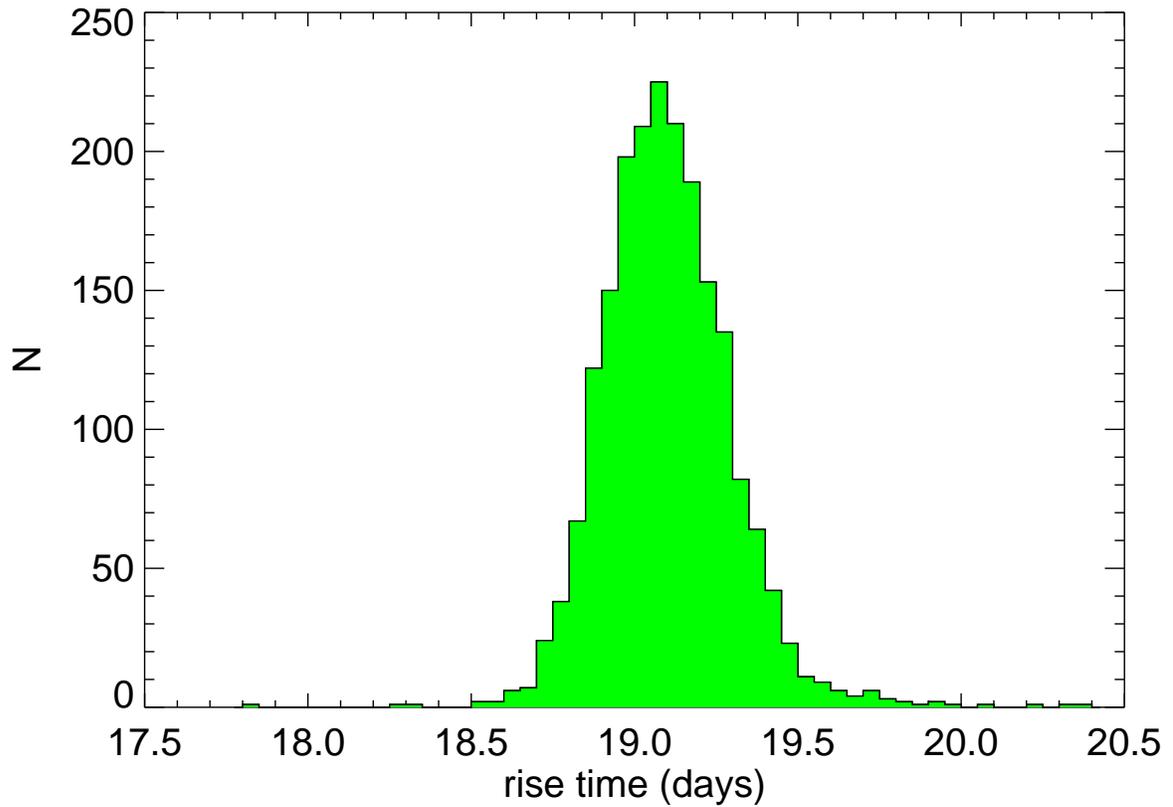}
\caption[\tr histogram for 73 SNLS SNe~Ia]
{Rise-time (\tr ) histogram for 73 SNLS SNe~Ia, taking into account
 the correlations between data points as described in 
 \S \ref{sec:parameterization}. This represents 2000 realizations of the light 
 curve parameters. The mean value is 19.10 days. \label{fig:maintrhisto}}
\end{figure}

\begin{figure}
\plotone{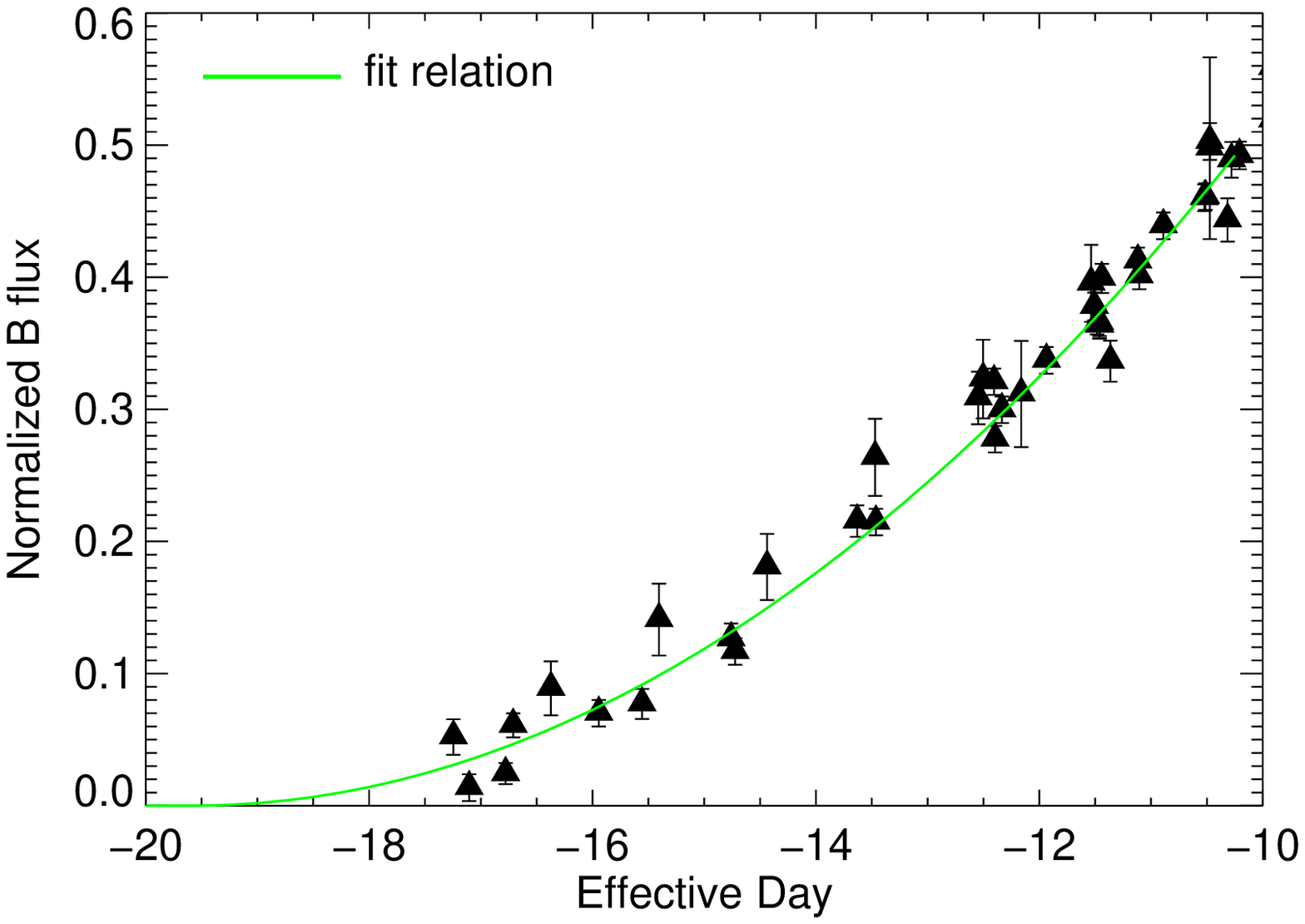}
\caption[Nearby rise-time data]{ Data in the rise-time
 region for the eight nearby SNe~Ia used in this study.
 These have been combined using the techniques described in
 \S \ref{sec:parameterization}. \label{fig:nearbyrisedata} }
\end{figure}

\nobreak
\label{sec:results}
For our sample of 73 high-redshift SNe~Ia, we find the rise time to be
$\tr = 19.10^{+0.18}_{-0.17}$ days (statistical errors only).
The histogram of rise-time values is shown in figure~\ref{fig:maintrhisto}.
This is approximately a factor of 6 more precise than the measurements 
of AKN00 or G01.  The low-redshift SN sample gives
$\tr = 19.58^{+0.22}_{-0.19}$. The nearby data in the rise-time region are
shown in figure~\ref{fig:nearbyrisedata}.  We resist comparing these until we
have estimated the systematic errors below.  The errors can be
roughly checked using a bootstrap analysis, which agrees with
those values quoted above.  Our estimates for the nuisance parameter
$1000 \times \alpha$  are $6.15^{+0.31}_{-0.31}$ for the SNLS sample and
$5.65^{+0.25}_{-0.27}$ for the nearby one.  $\alpha$ and \tr\ are quite 
correlated, with a correlation coefficient of $\rho = -0.6$.

In order to test the adequacy of the quadratic rise-time model, we
have also performed fits in which the exponent is allowed to vary in
the rise-time relation ($n$ in $f \propto t^n$).  The best constraint 
comes from low redshift, at which we find $n=1.7 \pm 0.2$, with the rise time 
reduced to $\tr = 18.80^{+0.37}_{-0.32}$ days.  Not surprisingly, the errors
on \tr\ are considerably larger if $n$ is not held fixed. The SNLS sample gives
$n=2.0^{+0.4}_{-0.3}$, with $\tr = 19.39^{+1.07}_{-0.82}$ days.
The combined value is $n=1.8 \pm 0.2$, essentially consistent with
the assumed value of $2$.  In order to test that our conclusions are
not too dependent on the value of $n$, we also re-fit the nearby
and distant samples using a fixed value of $n=1.8$, finding rise times of
$\tr = 18.97^{+0.19}_{-0.18}$ and $18.49^{+0.17}_{-0.15}$ days, 
respectively.  Fixing $n=1.8$ shifts the rise time but does not appreciably 
affect the difference between the two samples.  This conclusion also holds 
true when we compare subsets of the high-redshift sample, so we restrict
the discussion to $n=2$ subsequently.

\begin{figure}
\plotone{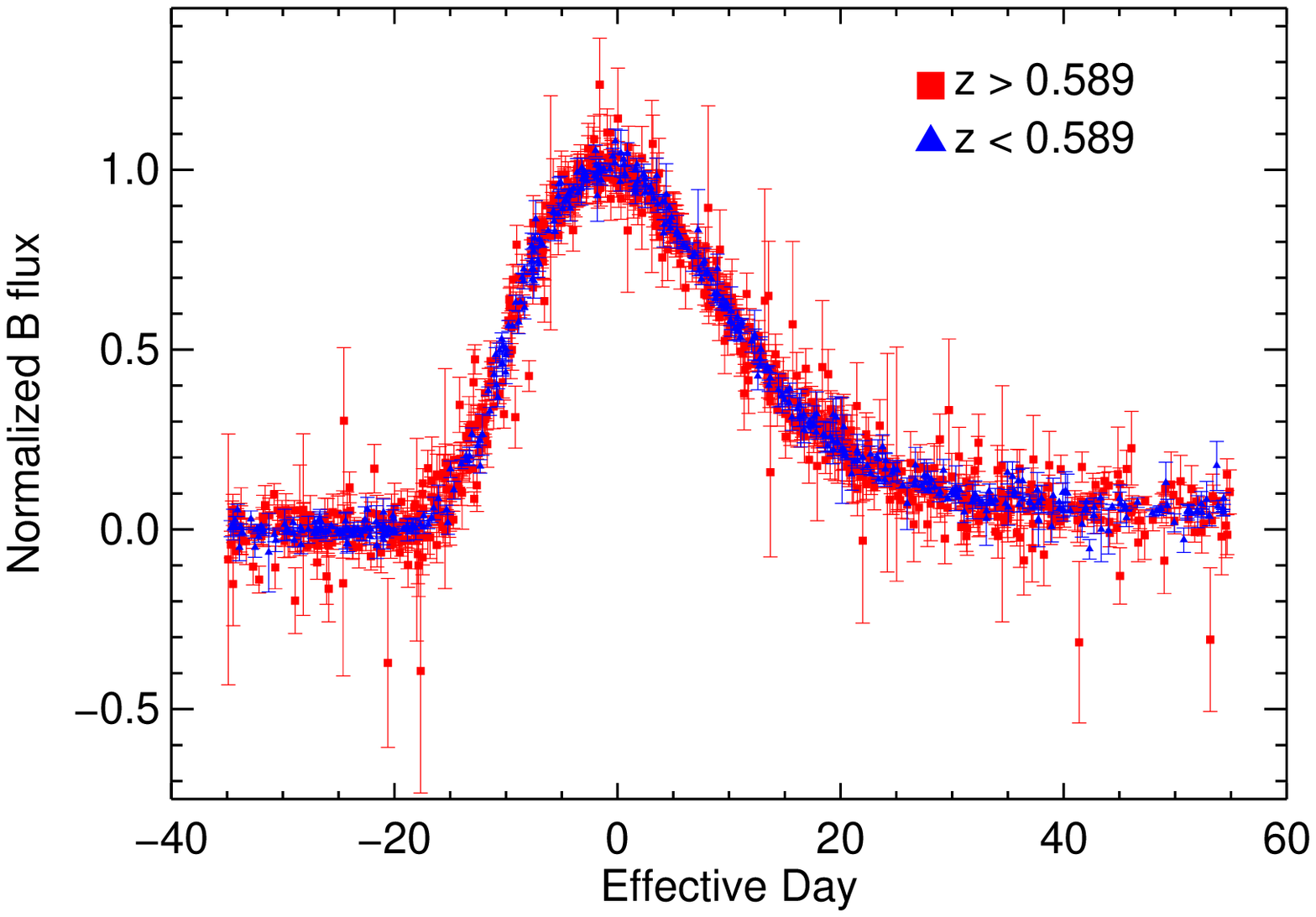}
\caption[Normalized and shifted SNLS data]
{ Shifted, normalized, and $K$-corrected SNLS data split by redshift
 and over-plotted.  The rest-frame $B$ band is shown. 
 Observations in the same filter on the same night are combined for
 display purposes; the actual fit is performed on the individual data
 points.  The blue triangles represent redshifts where rest-frame $B$
 matches the \rp\ filter, and the red squares where \ip\ is the best match.
 \label{fig:lcoverplot} }
\end{figure}

It is also quite interesting to compare the rise-time measurements for
different subsamples of our data. The results of these fits
are given in table~\ref{tbl:fitvals}.  We do not consider these subsets
of the nearby sample because it is too small.  First, we consider splitting
by redshift to search for evolution within our sample.  The median redshift
is 0.647.  This is quite close to the transition between observer
frame \rp\ matching with rest-frame $B$ and $\ip \mapsto B$, which takes
place at $z=0.589$.  This test is therefore sensitive to two 
possible effects: evolution, and a calibration mismatch between \rp\ and \ip .
Because we cannot test each of these independently, we split at $z=0.589$, 
which results in 29 SNe~Ia in the 
intermediate-$z$ sample and 44 in the high-$z$ sample 
($\left< z \right> = 0.43$ and $0.74$, respectively).  The photometric 
noise in the high-$z$ portion of the sample is much larger than in
the intermediate-$z$ portion, as shown in figure~\ref{fig:lcoverplot}.
This figure also demonstrates that data from different SNe~Ia can be
combined quite accurately using the techniques described in 
\S \ref{sec:parameterization}.

\begin{figure}
\plotone{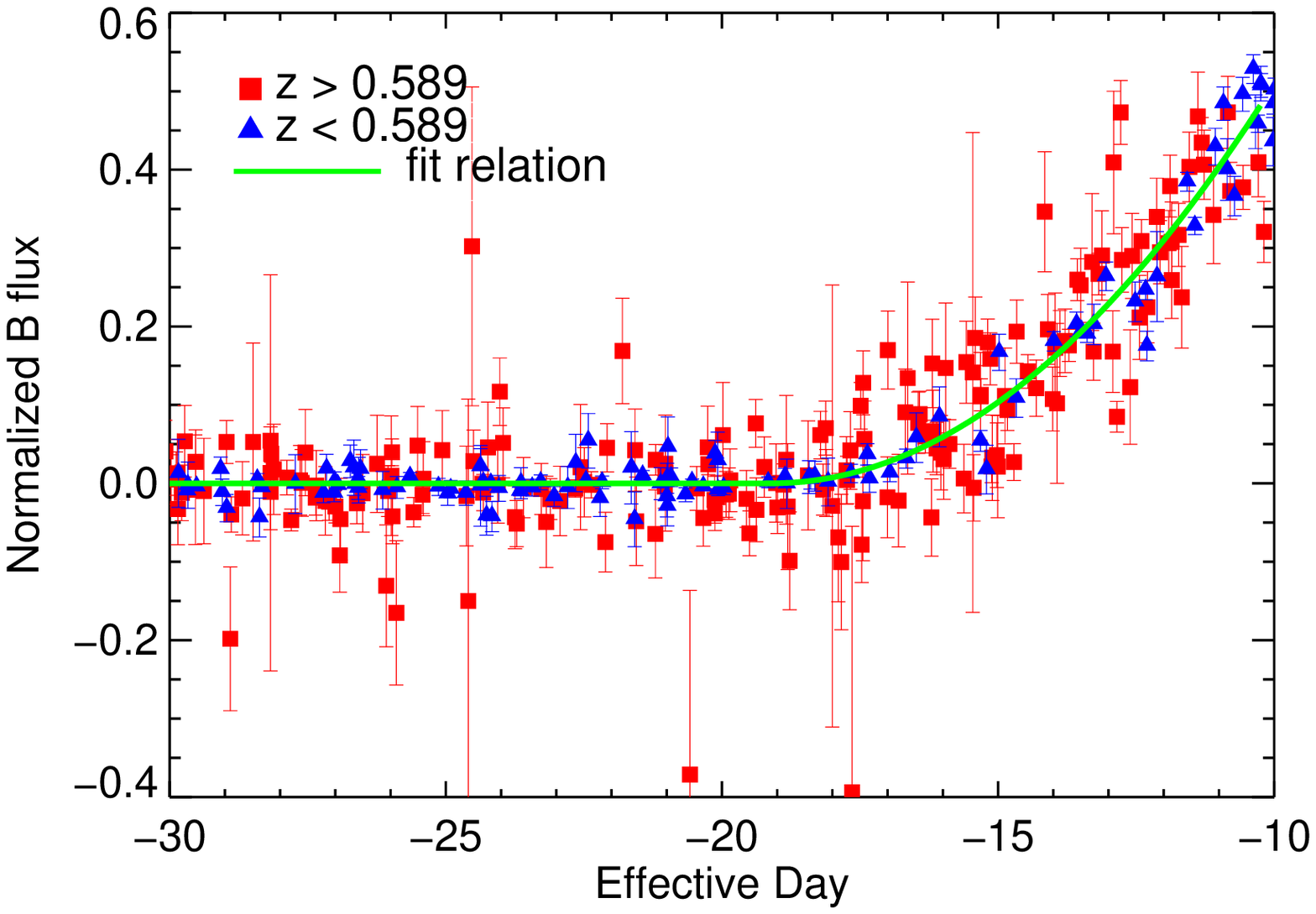}
\caption[SNLS rise-time data]{Data in the rise-time region for
 the 73 SNLS SNe~Ia used in this study, combined using the techniques
 described in \S \ref{sec:parameterization}.  They are 
 split by redshift as in figure~\ref{fig:lcoverplot}.  
 Observations from the same filter on the same night
 have been combined for display purposes. \label{fig:snlsrisetime} }
\end{figure}

The data in the rise-time region are shown in figure~\ref{fig:snlsrisetime},
again split into the two groups.  Using the same analysis, 
for $z \le 0.589$ we measure $\tr = 19.01^{+0.19}_{-0.18}$ days and for 
$z > 0.589$ we measure $\tr = 19.67^{+0.54}_{-0.49}$ days.  The 
intermediate-$z$ portion of the sample clearly dominates the fit to the 
full sample.  These are statistically compatible (the difference is 1.2 \sg ).

 As a test of whether the stretch model works at such early 
times, we split the sample by stretch.  Unlike the
nearby sample, one cannot measure the rise time precisely for
most of the distant SNe~Ia individually, so this test must be done on a 
sample-wide basis.  Splitting around the mean stretch, the 
low-stretch sample ($s \le 0.99, \left<s\right>=0.92$) of 36 SNe
gives $\tr = 19.20^{+0.33}_{-0.34}$ days and the high-stretch 
sample ($s > 0.99, \left<s\right>=1.05$, 37 SNe) gives 
$\tr = 19.09^{+0.33}_{-0.20}$ days.  These straddle the full sample 
value, and are statistically indistinguishable (0.3 \sg\ difference). 
If stretch did not work at early times, we generally would not expect 
these to agree.

\begin{figure}
\plotone{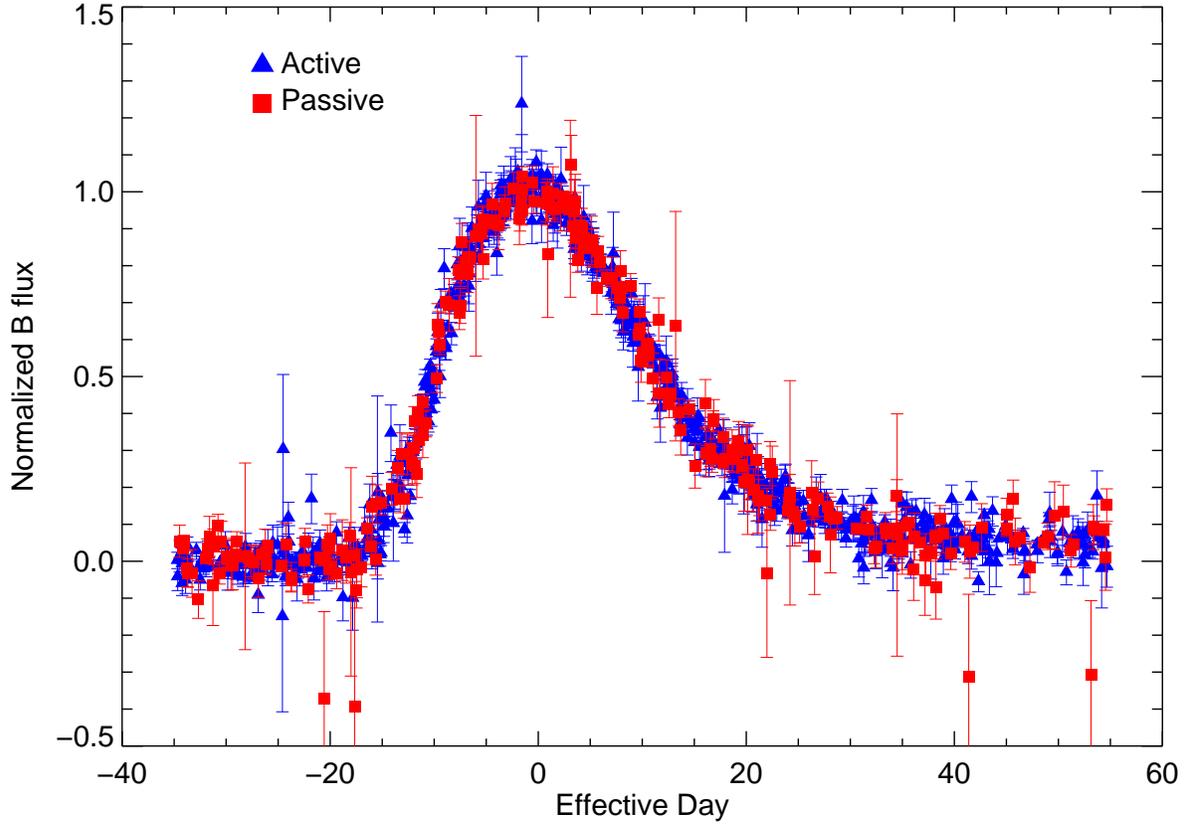}
\caption[SNLS rise-time data split by host galaxy type]
{ Shifted, normalized, and $K$-corrected SNLS data split by host
 galaxy specific star formation rate.  In all other respects this
 is identical to figure~\ref{fig:lcoverplot}.  The blue triangles 
 include SN from both active and vigorously star
 forming hosts as defined in \S \ref{sec:results} ($\log_{10}{sSFR} > -12$), 
 while the red squares show those from from passive hosts. \label{fig:lcsfr} }
\end{figure}

We can also split the sample by host galaxy star formation rate,
following the methodology of \citet{Sullivan:06}.  Briefly,
the broadband colors of the host galaxies are fitted by a galaxy
spectral evolution code using the known redshift to determine the
star formation rate per unit mass (sSFR), measured in yr$^{-1}$.  
The sample is then split between SNe in hosts with no star formation 
(passive, zero sSFR), those with moderate star formation rates 
(active, $-12 \le \log_{10}{sSFR} \le -9.5$), 
and those with a large amount of star formation (vigorously star forming,
$\log_{10}{sSFR} > -9.5$). We follow the above paper in limiting the
application of this technique to $z \le 0.75$, where it is most reliable
for our data set.  This results in a sample of 9 SNe~Ia in passive
hosts, 11 in active galaxies, and 35 in vigorously star forming
galaxies.  Clearly, this is an area where an increased sample size
would be beneficial.  Note that the comparison of these subsets is 
not independent of the stretch comparison: there is a relation 
between stretch and host galaxy morphology \citep{Hamuy:00} and, in addition,
star-formation rate \citep{Gallagher:05} for nearby SNe~Ia
that has recently been confirmed at high redshift in the SNLS 
sample \citep{Sullivan:06}.  In 
any case, for the passive sample we find a value of 
$\tr = 20.40^{+1.04}_{-1.10}$ days, for the active sample 
$\tr = 18.95^{+0.40}_{-0.41}$ days, and for the vigorously star-forming 
sample $\tr = 19.07^{+0.19}_{-0.18}$ days.  The rise time in passive hosts is
mildly different than the others but only at the 1.2 \sg\ level.  The data 
is shown in figure~\ref{fig:lcsfr}.

\begin{deluxetable}{lccl} 
\tablewidth{0pt}
\tablecaption{Results of rise-time fits}
\tablehead{ \colhead{ fit } & 
 \colhead{ $N_{\mbox{SN}}$ } & \colhead{ $\tr$ (days)\tablenotemark{a}} &
 \colhead{Notes}
}
\startdata
SNLS & 73 & $19.10^{+0.18\ 0.42}_{-0.17\ 0.36}$ & \nodata \\[3.5pt]
nearby & 8 & $19.58^{+0.22\ 0.46}_{-0.19\ 0.38}$ & \nodata \\[3.5pt]
SNLS intermediate $z$ & 29 & $19.01^{+0.18\ 0.37}_{-0.18\ 0.36}$ & 
 $z \le 0.589$, $\left< z \right> = 0.43$ \\[3.5pt]
SNLS high $z$ & 44 & $19.67^{+0.54\ 1.13}_{-0.49\ 0.99}$ & 
 $z > 0.589$, $\left<z\right> = 0.74$\\[3.5pt]
SNLS low $s$ & 36 & $19.20^{+0.33\ 0.67}_{-0.34\ 0.67}$ & 
 $s \le 0.99$, $\left<s\right> = 0.92$ \\[3.5pt]
SNLS high $s$ & 37 & $19.09^{+0.20\ 0.39}_{-0.20\ 0.40}$ & 
 $s > 0.99$, $\left<s\right> = 1.05$ \\[3.5pt]
SNLS passive hosts & 9 & $20.40^{+1.04\ 1.73}_{-1.10\ 2.06}$ 
 & $\log_{10}{sSFR} < -12.0$ \\[3.5pt]
SNLS active hosts & 11 & $18.95^{+0.40\ 0.74}_{-0.41\ 0.83}$ 
 & $-12 \le \log_{10}{sSFR} \le -9.5$ \\[3.5pt]
SNLS vigorous hosts & 35 & $19.07^{+0.19\ 0.40}_{-0.18\ 0.37}$ 
 & $\log_{10}{sSFR} > -9.5$ \\[3.5pt]
\enddata 
\tablenotetext{a}{Both the 68.3\% and 95.4\% confidence limits are given
 (1 and 2 \sg , respectively).}
\tablecomments{Rise-time fits to various samples.  sSFR is the star 
 formation rate per unit mass in units of $\mbox{yr}^{-1}$.}
\label{tbl:fitvals}
\end{deluxetable}

The dominant systematic error for our measurement should arise
from the \kcorrs .  Given the relative rarity of early-time
photometry of nearby SNe~Ia at early epochs, it is not surprising that there
is very little spectroscopy available in the rise-time region.  Hence,
our ability to combine data from SNe~Ia at different redshifts accurately
is highly dependent on the theoretical models used to derive the SED
at these epochs.  At sufficiently early times SNe~Ia should have SEDs
dominated by thermal continuum.  However, it is not observationally clear at 
what point this becomes a poor approximation.

We have tried to quantify this by considering the effects of
using a different, largely independently derived set of spectral templates,
specifically those of \citet{Nobili:03}.  Redoing the above analysis,
including the light-curve fits, shifts the rise time by 0.2 days.
We take this as an estimate of the systematic
uncertainty arising from \kcorrs .  AKN00 also considered the
effects of changing the late-time light-curve behavior, arguing that
this gave an upper limit for the systematic error of 2--3 days.  However,
the conditions considered were somewhat extreme, namely that {\it all}
of the high-redshift sample have unusual late-time behavior when compared
with the nearby sample.  As an upper limit this is reasonable,
but as a systematic estimate it is quite conservative.  We expect that
the systematic error involved in comparing different subsets of the SNLS
sample against each other should be smaller, since uncertainties in
the SED should affect the two samples in a similar fashion.  In any case,
since the subset comparisons reveal no effect even without systematic
errors, we have not tried to estimate them.

Ideally for the low-redshift sample \kcorrs\ should not be a problem. 
The SNe~Ia in the current nearby rise-time sample are all essentially at
zero redshift, and so if the observations were on the standard photometric
system there would essentially be no SED dependence of this process.
Unfortunately this is not the case; a significant portion of the
nearby data comes from unfiltered observations.  We are not in a position
to use the same technique to estimate the systematic error on the nearby
data.  R99 discussed some of the uncertainties associated with transforming
to the standard pass bands, but choose to include these effects as large
statistical errors rather than as an overall systematic effect.  Therefore,
we simply have to trust that our derived statistical errors incorporate
systematic uncertainties due to \kcorrs\ for the nearby sample.

\section{CONCLUSIONS}
\nobreak
We have measured the rise time from a sample of 73 high redshift
($z=0.15-0.9$) spectroscopically confirmed SNe~Ia discovered and
observed by the SNLS.  This determination is roughly 6 times
more precise than those previously available in this redshift range
(AKN00, G01).  Our measurement for this sample is 
$\tr = 19.10^{+0.18}_{-0.17} \left(\mbox{stat}\right) 
\pm 0.2 \left(\mbox{syst}\right)$ days.  Using the
same analysis technique on a sample of eight nearby SNe~Ia ($z < 0.1$),
we derive a value of $\tr = 19.58^{+0.22}_{-0.19}$ days, where the quoted error
incorporates both statistical and systematic errors.
These differ at the 1.4 \sg\ level.  In other words, using a considerably
more precise comparison made possible by a substantially better data set,
we find no compelling evidence for any difference between the rise times
of nearby and distant SNe~Ia.  It is important to understand the limitations 
of this measurement in terms of its constraints on theoretical models.  As 
was the case in R99, AKN00, and G01, the uncertainties presented
above are the error in the mean stretch-corrected rise times of the
two samples, {\it not} the scatter of rise times between individual SNe~Ia.
However, testing for differences between the two samples is still a very
useful check against evolutionary effects that may be affecting cosmological
analyses using SNe~Ia.

The above result suggests that we are currently limited by the
systematic uncertainty associated with \kcorrs\ in performing this
comparison.  Therefore, significant advances will likely require better
constraints on the early-time SEDs of SNe~Ia.  Alternatively, for a
large enough sample, it may be possible to constrain the rise-time
by only considering redshifts for which the observer and rest-frame filters
match particularly well, minimizing the \kcorr\ errors.

The quadratic rise-time model, motivated by simple physical arguments,
provides a good fit to the data.  Dropping this assumption, we 
find $n = 1.8 \pm 0.2$ for $f \propto t^n$ at early times.
Fitting for this extra parameter substantially weakens our
constraints on \tr , but does not indicate any discrepancies.  
Redoing the fits with $n$ fixed at this value
also does not significantly affect any of our results, simply decreasing
all of the measurements of \tr\ by $\sim 0.6$ days.

We have also split the SNLS sample into subsets
and searched for differences between them.  The comparisons
we consider are splitting the sample by redshift at $z=0.589$ to
test for evolution within our sample and for calibration systematics,
splitting by stretch, and splitting by host galaxy star formation rate.  
In all of these cases the subsamples give compatible rise times.  The 
fact that the low- and high-stretch samples agree confirms the
claim of G01 that the stretch parameterization works both at early and 
late times, at least up until around day 35.  Unlike the rise-time fit to the 
full sample, for many of these subsets we are limited by statistics, so 
as the survey continues more sensitive comparisons will be possible.
Once a full cosmological analysis on all these SN is available,
it will also be interesting to check whether there is any correlation between
the residual from the best fitting cosmology and rise time.

\acknowledgements
The authors would like to recognize the very significant
cultural role and reverence that the summit of Mauna Kea has
within the indigenous community of Hawai'i.  We are grateful for
our opportunity to conduct observations from this mountain.  We
would also like to thank Peter Nugent and Peter H\"{o}flich
for useful discussions.  We gratefully acknowledge the assistance of 
Pierre Martin and the CFHT Queued Service Observations team.  
Jean-Charles Cuillandre and Kanoa Washington were particularly 
indispensable in making possible real-time data reduction at CFHT.
Canadian collaboration members acknowledge support from NSERC and CIAR;
French members from CNRS/IN2P3, CNRS/INSU, and CEA.
This research has made use of the NASA/IPAC Extragalactic 
Database which is operated by the Jet Propulsion Laboratory, 
California Institute of Technology, under contract with the National 
Aeronautics and Space Administration.  The views expressed in this
article are those of the authors, and do not reflect the official position
of the United States Air Force, Department of Defense, or the U.S. 
Government.

\facility{CFHT (MegaCam)}

\mbox{~}
\end{document}